# Exploring the Use of ChatGPT as a Tool for Learning and Assessment in Undergraduate Computer Science Curriculum: Opportunities and Challenges

Basit Qureshi

Department of Computer Science, Prince Sultan University, Saudi Arabia

qureshi@psu.edu.sa

*Abstract*— The application of Artificial intelligence for teaching and learning in the academic sphere is a trending subject of interest in the computing education. ChatGPT, as an AI-based tool, provides various advantages, such as heightened student involvement, cooperation, accessibility and availability. This paper addresses the prospects and obstacles associated with utilizing ChatGPT as a tool for learning and assessment in undergraduate Computer Science curriculum in particular to teaching and learning fundamental programming courses. Students having completed the course work for a Data Structures and Algorithms (a sophomore level course) participated in this study. Two groups of students were given programming challenges to solve within a short period of time. The control group (group A) had access to text books and notes of programming courses, however no Internet access was provided. Group B students were given access to ChatGPT and were encouraged to use it to help solve the programming challenges. The challenge was conducted in a computer lab environment using Programming Contest Control (PC2) environment which is widely used in ACM International Collegiate Programming Contest (ICPC). Each team of students address the problem by writing executable code that satisfies certain number of test cases. Student teams were scored based on their performance in terms of number of successful passed testcases. Results show that students using ChatGPT had an advantage in terms of earned scores, however there were inconsistencies and inaccuracies in the submitted code consequently affecting the overall performance. After a thorough analysis, the paper's findings indicate that incorporating AI in higher education brings about various opportunities and challenges. Nonetheless, universities can efficiently manage these apprehensions by adopting a proactive and ethical stance towards the implementation of such tools.

*Keywords—ChatGPT, Academic assessment, programming concepts, Data Structures and Algorithms, Academic assessment.*

## I. Introduction

The progress made in machine learning and natural language processing (NLP) has led to a surge in research interest for chatbots, which have been employed in a diverse range of commercial and non-commercial applications [1-3]. The emergence of ChatGPT and other AI-powered text generation technologies has brought them into the spotlight for educators and scholars. The availability of ChatGPT and similar technologies has raised many questions for educational professionals, policymakers, and researchers [4-5]. Using ChatGPT in computer science education has the potential to offer several benefits.

In the context of informatics education, ChatGPT can be a valuable tool for teaching and learning fundamental concepts of computing such as programming languages, data structures, algorithms etc. It can help students to review and study, especially when they struggle to find study partners or don't have time to attend tutoring. The technology has the potential to streamline the research and learning process, making it easier for students to access information and complete assignments. Additionally, it can also grade and provide feedback on assignments, giving teachers more time to create engaging lesson plans and focus more attention on their students. However, it is important to control ChatGPT use to avoid potential negative impacts, especially for students. Teachers and lecturers are concerned that the chatbot may increase plagiarism and spread academic dishonesty, which could negatively impact academic integrity [3, 10]. While ChatGPT can provide a clear formula for producing written work, it may not necessarily improve critical thinking and problem-solving skills. Nevertheless, these potential dangers can be mitigated by providing students with tools and resources for identifying and avoiding inaccurate information.

Computer Science educators generally believe that hands-on experience in a computer lab environment, help students gain valuable experience [6-7]. In addition to effective book/classroom learning, practical work provides students with insight into the methods of science, research practice, and experimental methodology. The incorporation of scientific research into schools through pedagogical principles, rather than strict scientific formalism, is often categorized into three main learning goals: learning science, learning about science, and learning to do science, which involves engaging in and developing expertise in scientific inquiry and problem-solving [8]. Practical work has been shown to promote students' positive attitudes and enhance motivation for effective learning [9-10]. New approaches for teaching and assessing scientific inquiry and practices are essential for guiding students to make informed decisions.

Prince Sultan University's College of Computer and Information Sciences features three undergraduate programs, namely Computer Science, Information Systems, and Software Engineering, all of which are accredited by ABET. These programs mandate students to complete 124 credit hours of study [11]. Additionally, students who choose to enroll in any of these programs must take three consecutive courses - CS101 Programming I, CS102 Programming II, and CS210 Data Structures and Algorithms, during the first two years of study. These courses provide a comprehensive understanding of Java programming language concepts and structures. The CS101 course comprises a lab component in which students must attend weekly sessions to learn and apply problem-solving skills using a prevalent IDE. Conversely, CS102 and CS210 adopt a project-based approach to learning, where students collaborate on real-life scenarios and devise solutions to address them. The main emphasis of the CS102 project is to implement an object-oriented approach to storing data in a structured manner. The primary objective of the CS210 project is to utilize appropriate data structures and algorithms to solve the problem while adhering to specific time and space limitations. Once a student completes the CS210 course, he/she may take advanced courses in computing/engineering within their program of study.

This study aims to scrutinize the efficacy of ChatGPT in enhancing students' learning capabilities in the initial programming courses of a Computing curriculum. One of the principal objectives of this study is to gauge the effectiveness of ChatGPT in the students' learning process. Additionally, we seek to explore how effectively this technology can be employed to evaluate students' performance in online project assessments. With this objective in mind, students who were undertaking the CS210 - Data Structures and Algorithms course were motivated to take part in a computing challenge. During the programming challenge, two sets of students were given a limited time to solve programming problems. The control group (Group A) was granted access to textbooks and programming notes but was not given internet access. On the other hand, Group B was allowed to use ChatGPT to help tackle the programming problems. The challenge was conducted in a computer lab environment, employing the Programming Contest Control (PC2) system, which is widely used in the ACM International Collegiate Programming Contest (ICPC). Each team of students tackled the problem by crafting executable code that meets specific test case requirements. The performance of the student teams was graded based on the number of successful test cases. The outcomes demonstrate that students who used ChatGPT had an edge in terms of earned scores; however, inconsistencies and inaccuracies in the submitted code negatively impacted their overall performance.

We make the following contributions in this work

- In the experimental study, two sets of students participating in the challenge were observed, and their performance with and without ChatGPT was recorded, resulting in a dataset for further research.

- A comprehensive examination of the impact of ChatGPT on this challenge is provided, discussing the advantages and disadvantages of its use in education.

The paper is structured as follows: Section 2 contains a related works section that describes the ChatGPT technology and its recent applications in the education sector. Section 3 outlines the methodology used in this research, followed by section 4, which presents the results and outcomes. Section 5 provides a discussion analyzing the results and offering recommendations for educators and researchers. Finally, Section 6 concludes the paper.

II. RELATED WORKS

Although ChatGPT is a recent addition to the public domain, it has already captured significant research interest. In the following section, we present recent studies that describe the technology and explore its potential applications for research.

*A. What is ChatGPT?*

Machine learning, a subfield of artificial intelligence (AI), refers to the ability of computer systems to learn from experience without being explicitly programmed. With advancements in computing power, more data availability, and algorithmic improvements, deep learning has emerged as a high-performing predictive tool [1-4]. The field of Natural Language Processing (NLP) has witnessed significant growth over the past few years. However, the emergence of ChatGPT (Chat Generative Pre-trained Transformer) in November 2022 has generated a renewed interest and enthusiasm for this technology. ChatGPT, developed by OpenAI [17], is a large language model with remarkable abilities in comprehending and producing language that closely resembles human speech. Its impressive performance in answering questions, engaging in conversations, and generating coherent and contextually relevant responses has marked a significant milestone in the advancement of conversational Artificial Intelligence. Despite its advanced capabilities, ChatGPT is essentially a complex chatbot that can trace its roots in the early stages of Long Short-Term Memory (LSTM) development [14] and have been extensively employed in various fields, including automated customer service support, E-Commerce, Healthcare, and Education.

OpenAI's release of the GPT-3 model family has raised the bar considerably for its main rivals, Google and Facebook, and represents a significant milestone in the advancement of natural language processing (NLP) models [12]. The largest GPT-3 configuration, which consists of 175 billion parameters, 96 attention layers, and a batch size of 3.2 million training samples, was trained using 300 billion tokens, typically sub-words [15]. The training process of GPT-3 draws on the successful strategies employed by its predecessor, GPT-2, such as modified initialization, pre-normalization, and reverse tokenization. However, GPT-3 also introduces a new refinement based on alternating dense and sparse attention patterns. GPT-3 is designed as an autoregressive framework that can achieve task-agnostic goals through a few-shot learning paradigm [15]. With its ability to adapt to various tasks with minimal training data, it is a versatile and powerful tool for NLP applications.

All GPT (Generative Pre-trained Transformer) models, including the most recent GPT-4 model, are built based on the core technology of Transformers. The Transformer architecture was first introduced in the seminal paper "Attention is All You Need" by Vaswani et al. in 2017 [16], which has significantly impacted the deep learning research community. Transformers have revolutionized the processing of sequence-to-sequence models, surpassing traditional models based on recurrent neural networks by a significant margin. While Transformers follow the classical encoder-decoder architecture, they differ dramatically by integrating self-attention modules that excel in capturing long-term dependencies between input sequence elements (i.e., tokens). By leveraging this information, it efficiently determines the importance of each element in the input sequence. The self-attention mechanism computes a weight for each element based on its relevance to other tokens in the sequence. This enables Transformers to handle variable-length sequences more effectively and capture complex relationships between the sequence elements, thus improving performance on various natural language processing tasks. Trained on a vast dataset of 570 GB of Internet data, ChatGPT possesses several features that make it stand out among its peers, particularly in terms of accurate natural language generation [17]. These features are a significant advantage for ChatGPT.

*B. Applications of ChatGPT in education and research*

ChatGPT, a cutting-edge chatbot developed by OpenAI, has already captured the interest of researchers despite its recent release to the public domain. In this section, we will delve into recent studies that explore the numerous applications and research possibilities of ChatGPT.

In [13] Wang et.al., conducted an assessment of ChatGPT's ability to support various design, manufacturing, and engineering education tasks. Their findings highlight ChatGPT's impressive capacity to provide information, generate structured content, and propose initial solutions in a comprehensive and creative manner, demonstrating its potential to support tasks involving summarization, synthesis, and creation. They note that it struggles to comprehend questions accurately and lacks the ability to utilize knowledge effectively to generate appropriate solutions. In some instances, ChatGPT may even create non-existent rules or equations to produce solutions. Additionally, ChatGPT lacks a profound understanding of the underlying concepts of its responses, and its answers may not always be trustworthy, traceable, or verifiable.

In their study, Khan et al. [18] explored various medical education applications that could benefit from ChatGPT, such as creating personalized learning experiences and generating case studies. Additionally, the authors highlighted the potential of ChatGPT for clinical management purposes, such as documentation and decision support. Jianning Li et. Al., in [19] authored a systematic review on ChatGPT in Healthcare. They note that ChatGPT's interface is designed for QA, allowing it to be seamlessly integrated into existing clinical workflows, and providing real-time feedback. In addition to providing direct answers to specific questions, ChatGPT also offers justifications for its responses. Interestingly, these justifications and answers sometimes contain novel insights and perspectives, which could inspire new research ideas. Even open-ended questions can yield valuable information thanks to ChatGPT's unique capabilities. ChatGPT is not equipped to handle queries that require image interpretation. Unlike expert systems, ChatGPT lacks the ability to "reason" and the justifications it provides are a byproduct of its ability to predict the next word based on probabilities. The accuracy of ChatGPT's responses is heavily reliant on the quality of the data it was trained on, as the training data determines how ChatGPT responds to a given question. Kostick-Quenet and Gerke's in [20] emphasizes the importance of evaluating the real-world impact of ChatGPT and other language learning models (LLMs) to prevent any potential negative consequences resulting from their misuse. Although ChatGPT and other LLMs are innovative and revolutionary tools, their impact can be severely damaging if they are not used properly. Thus, it is crucial to assess their practical applications and potential risks before widespread implementation.

According to Fauzi et al. [21], ChatGPT can play a significant role in enhancing student productivity in higher education. The authors suggest that ChatGPT can provide valuable resources and information, improve language skills, promote collaboration, enhance time management, boost motivation, and increase efficiency and effectiveness. However, they also caution that ChatGPT should not be viewed as a replacement for human interaction or students' effort in achieving their academic goals. Instead, its role should be considered as a complement to these efforts. Kung et al. [22] conducted a study to determine if ChatGPT is capable of answering medical exam questions. They evaluated ChatGPT's performance on the US medical licensing exam, which includes three standardized tests that are mandatory for obtaining medical licensure in the US. The study excluded any questions containing visual information such as medical images. ChatGPT performed at a level near the passing threshold, achieving 60% accuracy without any specialized input from humans.

Pardos and Bhandari [23] conducted a study to determine the effectiveness of ChatGPT in providing hints to students in elementary and intermediate Algebra topics. They found that 70% of the hints offered by ChatGPT led to positive learning gains for students. Basic et al. [24] conducted a study with nine students in the control group and nine in the experimental group that used ChatGPT to determine whether it improves essay quality. Surprisingly, the control group outperformed the experimental group in most criteria, leading the authors to conclude that ChatGPT does not necessarily improve essay quality. In the field of software development, ChatGPT has been evaluated for its capabilities in tutoring and software testing. Jalil et al. [25] presented ChatGPT with textbook questions to evaluate its ability in software testing. Unfortunately, ChatGPT's performance in this case was not promising, as it was only able to provide correct answers to 37.5% of the questions.

In a nutshell, ChatGPT is a disruptive technology for its potential to fundamentally change the way academia and research are conducted. ChatGPT can generate personalized learning materials and case studies, improving the quality of education for students. However, there are also potential drawbacks to the use of ChatGPT in academia and research. There is a risk that the use of ChatGPT could lead to a loss of creativity and critical thinking skills, as researchers may rely too heavily on the tool's generated content. Additionally, there are concerns around the accuracy and reliability of ChatGPT's responses, particularly in cases where it may fabricate non-existing rules or equations.

Overall, ChatGPT represents a disruptive technology that has the potential to significantly alter the paradigm for academia and research. While there are both potential benefits and drawbacks to its use, it is clear that ChatGPT has already begun to have a significant impact on these fields and is likely to continue to do so in the years to come.

### III. METHODOLOGY

Research Question: Is there any statistical difference between the academic attainment level of students using ChatGPT to complete a programming challenge and the students who do not have the opportunity to do so.

Research Design: Quasi-experimental research [26] is typically carried out in real-world settings where it's either impossible or not feasible to randomly assign participants to groups. This type of research is commonly employed to assess the efficacy of a treatment or an educational intervention. To ensure the comparability of the control and experimental groups, all students who participated in the study were assessed with a pre-test to determine their proficiency in programming concepts. Additionally, all students had previously completed the three-programming course taught at Prince Sultan University CS101, CS102, and CS210 courses with a minimum grade of C+.

- CS101- Programming I is a 4 credit hours course that covers topics including Java Programming Language Syntax, Language Structure, Flow control, Loops, Methods, Arrays, File reading and writing.

- CS102 – Programming II is a 3 credit hours course covering topics such as Object-Oriented Programming in

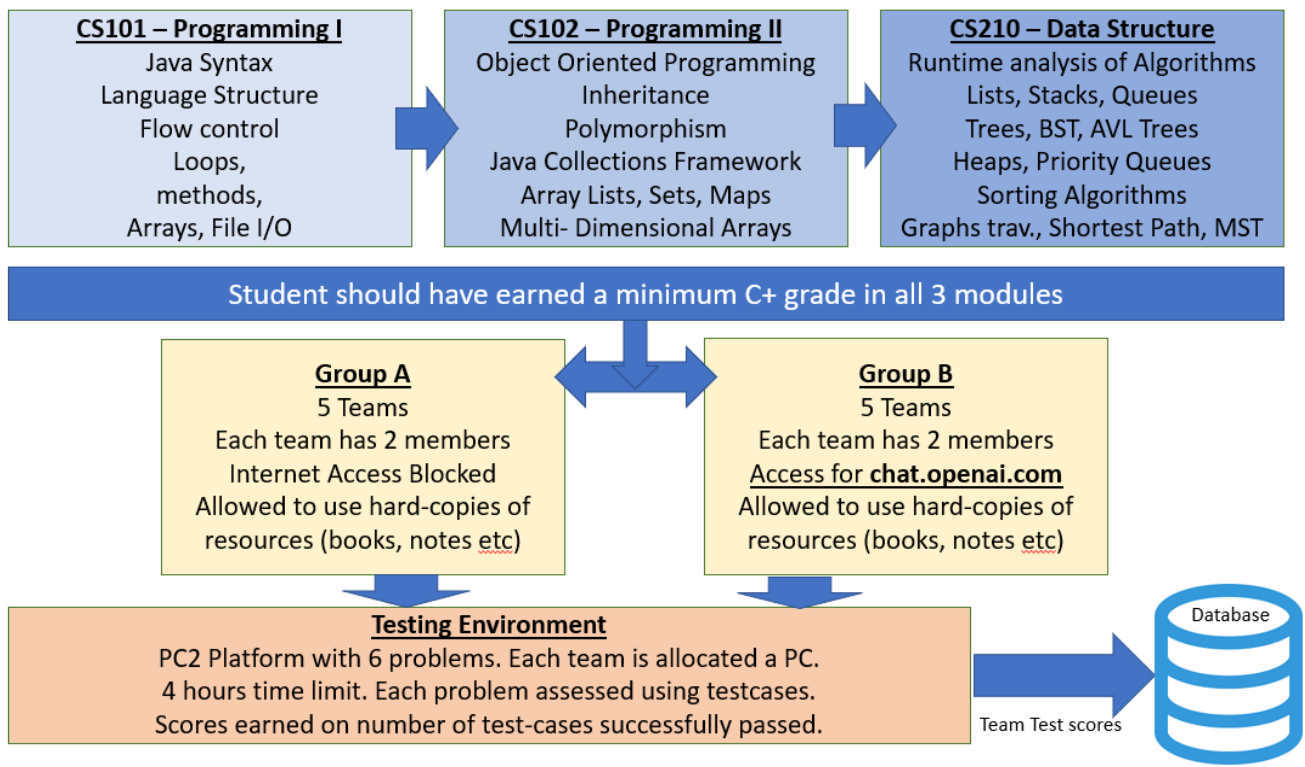

Figure 1: Overview of research methodology

Java, Inheritance, Polymorphism, Java Collections Framework, Array Lists, Sets, Maps, Multi- Dimensional Arrays.

• CS210 – Data Structure & Algorithms is a 3 credit hours course that covers Runtime analysis of Algorithms, Lists, Stacks, Queues, Trees, BST, AVL Trees, Heaps, Priority Queues, Sorting Algorithms, Graphs traversal algorithms, Shortest Path, and Minimum Spanning Trees.

After finishing the pre-test, the students were separated into two groups: a control group (group A) and an experimental group (group B). Each group was composed of six teams, and each team had two members. The study involved a total of 24 students. The purpose of this was to ensure that both groups were homogeneous with respect to academic level and pre-test scores. The intervention or challenge was then administered, and the post-test scores of the students were used to measure their achievement levels. The resulting data was collected and statistically analyzed to identify any significant differences in the mean scores between the control and experimental groups. Figure 1 shows an overview of the research design.

Instrumentation: The challenge is specifically crafted to evaluate the participants' algorithmic and programming capabilities, as well as their problem-solving skills. During the competition, teams are presented with a series of six programming problems, each assigned a point value based on its difficulty level. Teams earn points for every problem they solve correctly within the stipulated time limit. Notably, each problem in the challenge is accompanied by a set of hidden test-cases, with the number of earned points depending on the number of test-cases a student-team can successfully complete. The challenge comprises six programming problems, each with its own statement outlining its requirements, input/output conditions, and limitations. To encourage efficient problem-solving, each problem has prescribed memory and CPU time limitations, thereby motivating students to use optimized data structures in their solutions.

The challenge was conducted in a computer lab environment using Programming Contest Control (PC2) [32] environment which is widely used in ACM International Collegiate Programming Contest (ICPC). Each team of students comprising of 2 students addressed the challenge-problems by writing executable code that satisfies certain number of test cases. Student teams were scored based on their performance in terms of number of successful passed testcases. Time taken by all teams for each problem was also recorded for statistical evaluation.

TABLE 1: SET OF PROBLEMS USED

| Prob. No | Data Structure / Algorithm | # of Test Cases | Time, Memory Req. |
|---|---|---|---|
| P1 | Lists traversal | 10 | $O(n)$, $O(n)$ |
| P2 | AVL trees | 8 | $O(\log n)$, $O(n)$ |
| P3 | Sorting Arrays | 14 | $O(n \log n)$, $O(n)$ |
| P4 | Graph topological sorting | 8 | $O(V+E)$, $O(V+E)$ |
| P5 | Shortest Path / Dijkstra Algo | 12 | $O((V+E) \log V)$, $O(V)$ |
| P6 | MST using Prim's algorithm | 13 | $O(E \log V)$, $O(V+E)$ |

Table 1 shows the data-structure used for each of the six problems, with the number of test-cases defined in the system. The time and memory requirements are also given. Here problems P1-P3 are considered easy, P4-6 increase in difficulty.

Data collection process: The control panel in PC2 provides a detailed view on students scores earned, time spend in each problem, the number of times a student submitted their work, number of test-cases passed etc. In this study, the results of interest were the team scores earned and the time taken by a team to submit their work.

TABLE 2: PERFORMANCE OF GROUP A STUDENT TEAMS

| P# | A1 | A2 | A3 | A4 | A5 | A6 |
|---|---|---|---|---|---|---|
| P1 | 100 | 80 | 100 | 90 | 70 | 100 |
| P2 | 87.5 | 62.5 | 75 | 50 | 62.5 | 100 |
| P3 | 78.5 | 42.8 | 64.3 | 57.1 | 71.4 | 85 |
| P4 | 62.5 | 37.5 | 50 | 25 | 25 | 75 |
| P5 | 58.3 | 25 | 42.6 | 16.6 | 8.33 | 83.3 |
| P6 | 53.8 | 30.7 | 7.69 | 0 | 0 | 38.4 |
| *Total* | *440.60* | *278.50* | *339.59* | *238.70* | *237.23* | *481.70* |
| *Max* | *100.00* | *80.00* | *100.00* | *90.00* | *71.40* | *100.00* |
| *Min* | *53.80* | *25.00* | *7.69* | *0.00* | *0.00* | *38.40* |
| *Avg* | *73.43* | *46.42* | *56.60* | *39.78* | *39.54* | *80.28* |
| *Std Dev* | *18.24* | *20.90* | *31.35* | *32.45* | *32.31* | *22.76* |

TABLE 3: PERFORMANCE OF GROUP B STUDENT TEAMS USING CHATGPT

| P# | B1 | B2 | B3 | B4 | B5 | B6 |
|---|---|---|---|---|---|---|
| P1 | 100 | 70 | 100 | 100 | 100 | 100 |
| P2 | 50 | 62.5 | 87.5 | 100 | 100 | 87.5 |
| P3 | 100 | 71.4 | 100 | 64.3 | 75 | 100 |
| P4 | 75 | 0 | 87.5 | 75 | 87.5 | 87.5 |
| P5 | 83.3 | 16.6 | 0 | 66.6 | 58.3 | 91.6 |
| P6 | 69.2 | 0 | 7.69 | 38.4 | 61.53 | 76.9 |
| *Total* | *477.50* | *220.50* | *382.69* | *444.30* | *482.33* | *543.50* |
| *Max* | *100.00* | *71.40* | *100.00* | *100.00* | *100.00* | *100.00* |
| *Min* | *50.00* | *0.00* | *0.00* | *38.40* | *58.30* | *76.90* |
| *Avg* | *79.58* | *36.75* | *63.78* | *74.05* | *80.39* | *90.58* |
| *Std Dev* | *19.24* | *34.86* | *46.83* | *23.53* | *18.40* | *8.77* |

TABLE 4: MEANS AND STANDARD DEVIATIONS FOR P1-P6 FOR GROUPS A AND B

| Problems | Group A | | Group B | |
|---|---|---|---|---|
| | **MEAN** | **STD** | **MEAN** | **STD** |
| **P1** | 95.00 | 12.65 | 100.00 | 12.25 |
| **P2** | 68.75 | 18.40 | 87.50 | 20.54 |
| **P3** | 67.85 | 15.27 | 87.50 | 16.66 |
| **P4** | 43.75 | 20.41 | 81.25 | 34.23 |
| **P5** | 33.80 | 28.24 | 62.45 | 36.75 |
| **P6** | 19.20 | 22.49 | 49.97 | 32.54 |

## IV. RESULTS

Table 4 shows the mean and SD for scores earned by groups A and B for problems P1 to P6. It is a common practice to breakdown numerical data in order to provide a visual representation that can highlight any variations in scores and deviations. The mean score for Group B is significantly higher than Group A for all problems. Both groups had fairly similar SDs for Problems P1, P2 and P3, however the difference is significant for P4 – P6.

Figure 2 shows the average team scores for group A and B for the challenge. Overall Group B teams score better compared to group A teams.

Figure 3 shows the time taken by teams to complete the challenges P1 to P6. Group A teams took more time to complete challenges P1 to P4. However, Group B teams took more time to complete challenges P5 and P6. Group A teams were not able to score 100% in any of P3 to P6 problems. Significant amount of time was used in debugging and fixing errors in problems P3 and P4, consequently poor time management resulted in low scores earned in P3-P6. Teams B1 to B6 used ChatGPT to help write code.

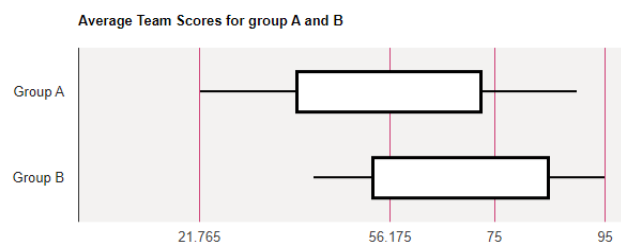

Figure 2: Average team scores for group A and B

Observations were made that indicated teams B2 to B5 utilized code generated by ChatGPT, but faced challenges in passing all the test cases within the PC2 environment. This was due to the fact that some of the generated code was too lengthy and exceeded ChatGPT's output size limits, which resulted in truncation. As a result, more time was spent debugging the code to resolve these issues. Additionally, multiple queries were made to ChatGPT, leading to inconsistencies. On the other hand, teams B1 and B6 also used ChatGPT to develop their code, but they did not solely rely on the generated code. They employed the suggested algorithms to write their programs. However, none of the Group B teams were able to score 100% on problems P4 to P6. As a result, the majority of teams in group B spent time on resolving errors by debugging their code.

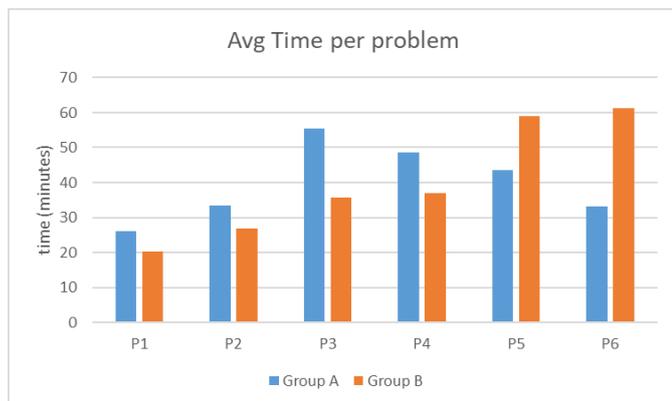

Figure 3: Time taken by teams to complete the challenge

## V. DISCUSSION

Our findings suggest that ChatGPT is remarkable in producing coherent and well-structured code, and proposing optimized solutions against the problem statement. As a generative model, ChatGPT generates promising outputs and a comprehensive explanation for students to address the given problem. If prompted correctly, it gives step by step guidance explaining the underlying algorithm to be used in the solution with code snippets to help understanding. Students prompted it to write complete code for the given problem, which it did, however executing this code in the IDE gave mixed results.

We also identified limitations of ChatGPT in writing correct code/program. It faces challenges in comprehending questions and lacks the proficiency to utilize knowledge to generate accurate solutions. In some cases, the generated code snippet did not compile using the IDE after several prompts to fix the code, causing frustration. Additionally, inaccuracies were observed with the output of the generated code which led in failing the test-cases. This was particularly noticeable as the problems difficulty level increased for the students (P4 onwards). In addition, because of the restrictions on the length of the generated text, it was occasionally impossible to obtain the entire program. As a result, students had to prompt the ChatGPT multiple times to acquire the desired output, which led to compile errors.

For a student, having a thorough grasp of the inner workings of the algorithms is unnecessary, but comprehending the variables that will influence the result is crucial. It is important to inquire about the corpus of data that was utilized to train the AI, including any omissions and selection criteria. Additionally, it is important to consider the possible implicit biases and assumptions that the tool may have.

To generate the most useful content, students must be proficient in providing precise prompts to the AI bot. In this sense, large language models are comparable to search engines since they require appropriate prompts to yield relevant results. Crafting effective prompts necessitates a deep understanding of the tool and the content that underlies the prompt, as well as critical thinking. While a basic prompt may suffice in producing the desired output, significant effort is required to obtain the desired outcome.

Students from Group B were urged to copy/save prompts used to generate code. Some prompts used by students:

- Write java code to address this problem (problem statement).
- This code did not compile, write a better code.
- This code still did not compile, fix the errors.
- This error was generated, fix this error and re-write code.
- Your code works but it does not pass testcases.
- This code is too slow, write a faster program.
- This code runs out of memory. Write a better program that takes O(n) time.
- This program fails this test case. Fix the program so that it passes (test case input with desired output pasted at prompt).

The use of AI applications such as ChatGPT directed towards students has tremendous potential in enhancing intelligent student support systems and individualized learning. ChatGPT serves as an Intelligent tutoring system (ITS) that can help provide customized instructions to enhance learning process. Figure 4 shows an example of tailored learning when the ChatGPT is prompted: "Visually explain how an AVL tree attains O(log n) runtime?"

As education researchers continue to explore new learning models, opportunities in the field are expanding, and personalization is growing. With the aid of AI algorithms, ITSs can now imitate the support provided by a tutor, such as offering tailored assistance in problem-solving. It is expected that the advancements in big data technology and learning analytics will lead to a groundbreaking paradigm of adaptive, individualized learning.

Typically, when innovative educational technologies are introduced in the classroom [27-31], it can pose several challenges to the traditional teaching and learning methods. Managing these challenges becomes the responsibility of education practitioners and policymakers [33]. If left unaddressed, these challenges can potentially reveal deficiencies in pedagogical practices. In the context of teaching and learning programming in a computing curriculum, it is essential to see students learn by doing and gain experience writing code. Our findings suggest that students can easily generate working code in any programming language of their choice using ChatGPT. In most cases, the code generated, executes successfully after several prompts. Eventually the technology would catch up producing better results in fewer prompts.

To prevent misuse of ChatGPT as a tool for learning programming, there are several strategies that can be employed:

- Clear expectations: Educators should communicate clearly to students that ChatGPT is not a suitable resource for learning programming and should not be used for that purpose.
- Encourage hands-on learning: Emphasize the importance of hands-on learning and encourage students to practice

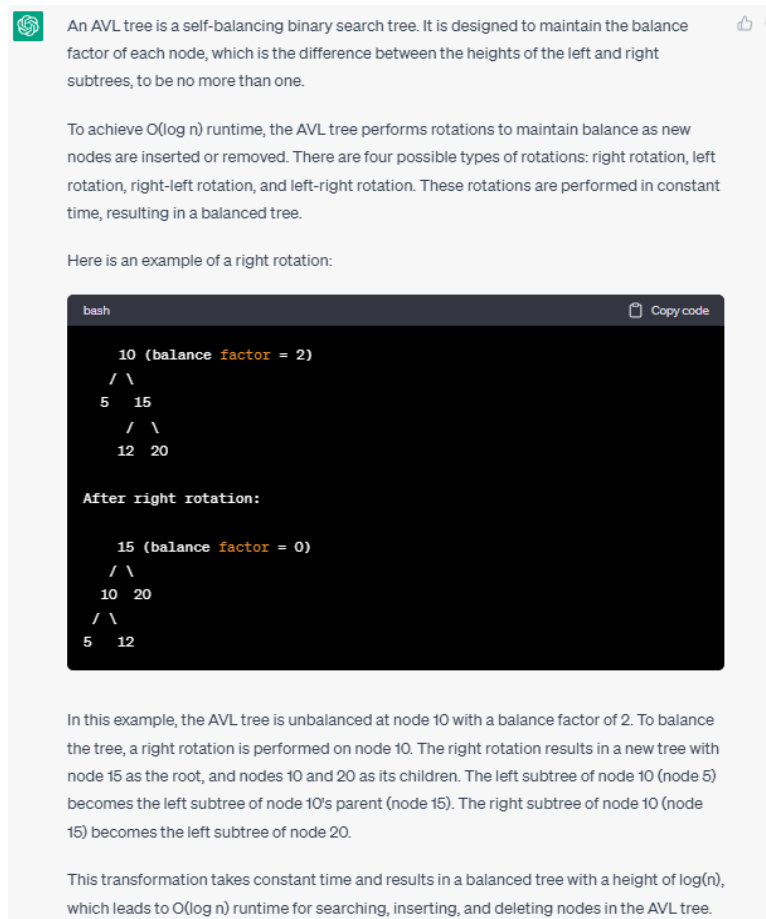

Figure 4: A sample of tailored response to a query

programming on their own using legitimate programming resources, such as textbooks, online courses, and coding platforms.

- Monitor usage: Instructors need to be vigilant and keep an eye on how students are using ChatGPT and intervene if there is evidence that it is being used inappropriately. This can be done by monitoring students' online activity or by checking their programming work for signs of plagiarism or excessive reliance on external resources.

- Assess learning outcomes: Incorporate assessments that test students' ability to apply programming concepts and skills, rather than just regurgitating information. This will encourage students to engage in active learning and discourage them from relying on ChatGPT for answers.

- Plagiarism detection tools: Moss (Measure of Software Similarity) is a plagiarism detection tool developed by Stanford University. It is designed to detect similarities between programming assignments submitted by students, and it can be used to help instructors identify cases of potential plagiarism. There are several other tools that can be used to getect plagiarism including JPlag, Codequiry etc.

- Interviews can be a useful tool for detecting plagiarism in cases where software tools may not be effective, such as when the student has modified code to make it appear original. However, this method requires significant time and effort on the part of the instructor and may not be feasible for large classes or assignments with many students.

## VI. CONCLUSIONS

The integration of AI in higher education teaching, learning, and assessment, presents both opportunities and challenges. An experimental study was conducted to assess the effectiveness of using ChatGPT for solving programming problems. The study involved 24 students who were divided into two groups control and experimental. According to the findings, the group that utilized ChatGPT in the experiment performed better than the control group, achieving higher scores in less time. However, the experimental group was not able to achieve perfect scores due to inaccuracies or inconsistencies in the code generated by ChatGPT.

Overall, it is important to create a learning environment that emphasizes the importance of active, hands-on learning and discourages students from relying on external resources such as ChatGPT for their programming education. As AI tools for solving programming challenges continue to emerge and improve in accuracy, students may be tempted to cheat if they do not fully understand the value of their learning. It is the responsibility of educators to clearly explain the significance of their work and the benefits of learning through practice. At the institutional level, academic integrity policies and honor codes should be revised to address the use of AI tools. Clear and simple guidelines for the proper use of language models in teaching and learning should be developed, and consequences for cheating should be clearly outlined. Faculty should be trained to adopt AI tools to augment their teaching practices. Additionally, students should receive training on academic integrity to ensure they fully understand the importance of maintaining ethical standards in their work.

ACKNOWLEDGMENT

The author would like to acknowledge the role of ChatGPT in the development of this paper. The author utilized the tool/service to enhance readability of his/her own sentences, and summarize and clarify text from related works section. The author takes full responsibility for reviewing and editing the content as necessary prior to publication.